\documentclass[12pt,draftclsnofoot,onecolumn,journal,letterpaper]{IEEEtran}
\usepackage[utf8]{inputenc} 
\usepackage[T1]{fontenc}
\usepackage{url}
\usepackage{ifthen}
\usepackage[cmex10]{amsmath}
\usepackage{tablefootnote}
\usepackage{array} 

\usepackage{enumerate}
\usepackage{amssymb}
\usepackage{amsmath} 

\usepackage{algorithm}
\usepackage{algorithmicx}
\usepackage{algpseudocode}

\usepackage{mathrsfs}
\usepackage{bm}
\usepackage{tikz}
\usetikzlibrary{arrows}
\usepackage{subfigure}
\usepackage{graphicx,booktabs,multirow}
\usepackage{epstopdf}
\usepackage{subfigure}
\usepackage{multirow}
\usepackage{booktabs}
\usepackage{makecell}
\usepackage{tabularx} 
\usepackage{booktabs}
\newcommand{\vct}[1]{{\bm #1}}
\newcommand{\mtx}[1]{{\bm #1}}
\DeclareMathOperator*{\argmin}{arg\,min}

\definecolor{colorhkust}{RGB}{20,43,140}
\definecolor{colortsinghua}{RGB}{116,52,129}
\definecolor{color1}{RGB}{128,0,0}

\usepackage{amsthm}
  
\newtheorem{thm}{Theorem}

\theoremstyle{definition}

\theoremstyle{remark}

\begin{document}

        \title{LORM: Learning to Optimize for Resource Management in Wireless Networks with Few Training Samples}

   \author{Yifei Shen, \textit{Student Member}, \textit{IEEE}, Yuanming Shi, \textit{Member}, \textit{IEEE}, Jun Zhang, \textit{Senior Member}, \textit{IEEE}, and Khaled B. Letaief, \textit{Fellow}, \textit{IEEE}
        
        \thanks{The materials in this paper were presented in part at the IEEE Global Conference on Signal and Information Processing (GlobalSIP), 2018 \cite{shen2018scalable}, and IEEE International Conference on Communications (ICC), 2019 \cite{shen2018transfer}.
        	
        	Y. Shen and K. B. Letaief are with the Department of Electronic and Computer Engineering, Hong Kong University of Science and Technology, Hong Kong (E-mail: \{yshenaw, eekhaled\}@ust.hk). Y. Shi is with the School of Information Science and Technology, ShanghaiTech University, Shanghai 201210, China (E-mail:  shiym@shanghaitech.edu.cn). J. Zhang is with the Department of Electronic and Information Engineering, The Hong Kong Polytechnic University, Hong Kong (E-mail: jun-eie.zhang@polyu.edu.hk). (The corresponding author is J. Zhang.)}
        
    }
        
        \maketitle

\begin{abstract}
Effective resource management plays a pivotal role in wireless networks, which, unfortunately, results in challenging mixed-integer nonlinear programming (MINLP) problems in most cases. Machine learning-based methods have recently emerged as a disruptive way to obtain near-optimal performance for MINLPs with affordable computational complexity. There have been some attempts in applying such methods to resource management in wireless networks, but these attempts require huge amounts of training samples and lack the capability to handle constrained problems. Furthermore, they suffer from severe performance deterioration when the network parameters change, which commonly happens and is referred to as the \emph{task mismatch} problem. In this paper, to reduce the sample complexity and address the feasibility issue, we propose a framework of Learning to Optimize for Resource Management (LORM). Instead of the end-to-end learning approach adopted in previous studies, LORM learns the optimal pruning policy in the branch-and-bound algorithm for MINLPs via a sample-efficient method, namely, \emph{imitation learning}. To further address the task mismatch problem, we develop a transfer learning method via self-imitation in LORM, named \emph{LORM-TL}, which can quickly adapt a pre-trained machine learning model to the new task with only a few additional \emph{unlabeled} training samples. Numerical simulations will demonstrate that LORM outperforms specialized state-of-the-art algorithms and achieves near-optimal performance, while achieving significant speedup compared with the branch-and-bound algorithm. Moreover, LORM-TL, by relying on a few unlabeled samples, achieves comparable performance with the model trained from scratch with sufficient labeled samples.

\end{abstract}

\section{Introduction}
\subsection{Motivations}
In wireless networks, effective management of radio resources is vital for performance optimization \cite{han2008resource}. Unfortunately, typical resource management problems, such as subcarrier allocation in OFDMA \cite{wong1999multiuser}, user association \cite{ye2013user}, access point selection \cite{shi2014group}, and computation offloading \cite{mao2016dynamic}, are mixed integer nonlinear programming (MINLP) problems, which are NP-hard. The complexity of global optimization algorithms, e.g., the branch-and-bound algorithm, is exponential. Thus, most of the existing studies focused on sub-optimal or heuristic algorithms, whose performance gaps to the optimal solution are difficult to quantify and control.

Machine learning has recently emerged as a disruptive technology to balance the computational complexity and the performance gap for solving NP-hard problems, and has attracted lots of attention from the mathematical optimization community \cite{yoshua2018machine}. This trend has also inspired researchers to apply machine learning-based methods to solve optimization problems in wireless networks \cite{zappone2018model}. For example, to solve the power control problem in the interference channel, it was proposed in \cite{sun2018learning} to use a neural network to approximate the classic weighted minimum mean square error (WMMSE) algorithm to speed up the computation. However, as the training samples are obtained via a sub-optimal algorithm, i.e., WMMSE, there is a performance gap compared with the optimal solution. To achieve near-optimal performance, unsupervised learning methods have been proposed in \cite{liang2018towards,lee2018deep}, which do not rely on any existing resource management algorithm. Besides improving performance and computational efficiency, machine learning techniques have also been applied to other problems in resource management. In particular, deep reinforcement learning \cite{nasir2018deep}, spatial deep learning \cite{cui2018spatial}, and deep neural network parameterization \cite{eisen2018learning} have been proposed to deal with the scenarios with delayed channel state information (CSI), without CSI but only geographical locations, and with unknown resource allocation functions, respectively.

While the attempts \cite{sun2018learning,liang2018towards,lee2018deep} have demonstrated the great potential of the ``learning to optimize'' approach for resource management in wireless networks, applying them into real systems faces additional difficulties. A prominent shortcoming of these methods is that they require large amounts of training problem instances, e.g., millions of samples are needed for a small-size system \cite{liang2018towards}. This incurs a significant cost for sample acquisition, and may not be feasible in practice. Secondly,
resource management problems are constrained by nature, but the ability of existing machine learning-based methods in dealing with constraints is limited \cite{yoshua2018machine}. Finally, wireless networks are inherently dynamic, e.g., both the locations and number of users are changing dynamically. Thus, a pre-trained machine learning model may be useless or suffer from severe performance deterioration as the network setting changes. This challenge can be characterized as \emph{task mismatch}, i.e., the test setting is different from the trained one, and it has not been treated in previous studies. To address these challenges, in this paper we develop a novel machine learning framework for MINLP resource management problems in wireless networks, which only requires a small number of training problem instances, achieves near-optimal performance with feasibility guarantee, and is able to effectively handle task mismatch.

\subsection{Literature Review}

\begin{figure}[htbp]
	\centering
	\subfigure[End-to-end learning, where the neural network takes the problem data as input and outputs the solution.]{
		\begin{minipage}[t]{0.47\linewidth}
			\centering
			\includegraphics[width=0.7\linewidth]{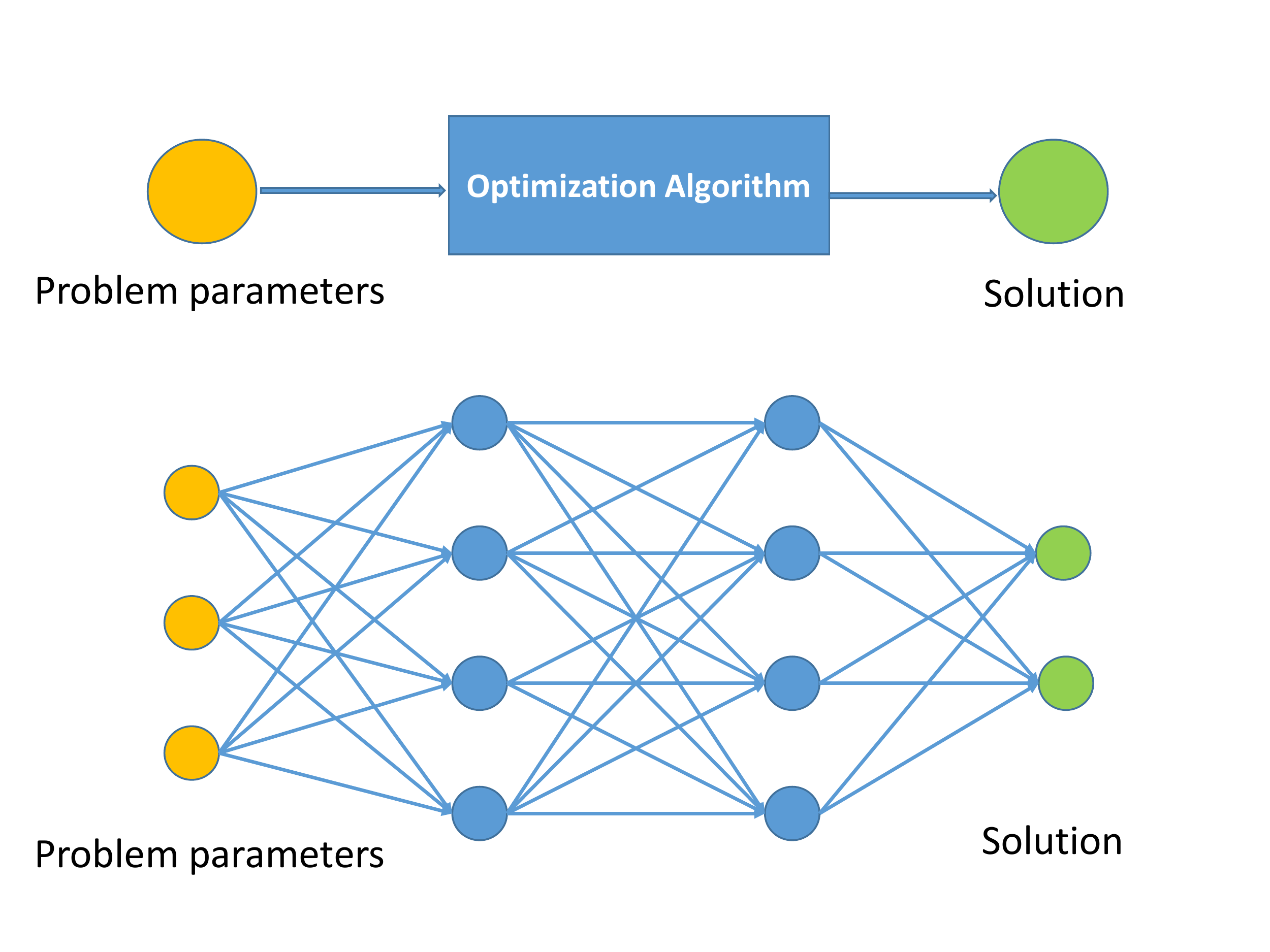}
			\label{fig:e2e} 
		\end{minipage}%
	}%
	\hspace{.1in}
	\subfigure[Optimization policy learning, where the neural network learns the optimal policy in the search tree.]{
		\begin{minipage}[t]{0.47\linewidth}
			\centering
			\includegraphics[width=0.7\linewidth]{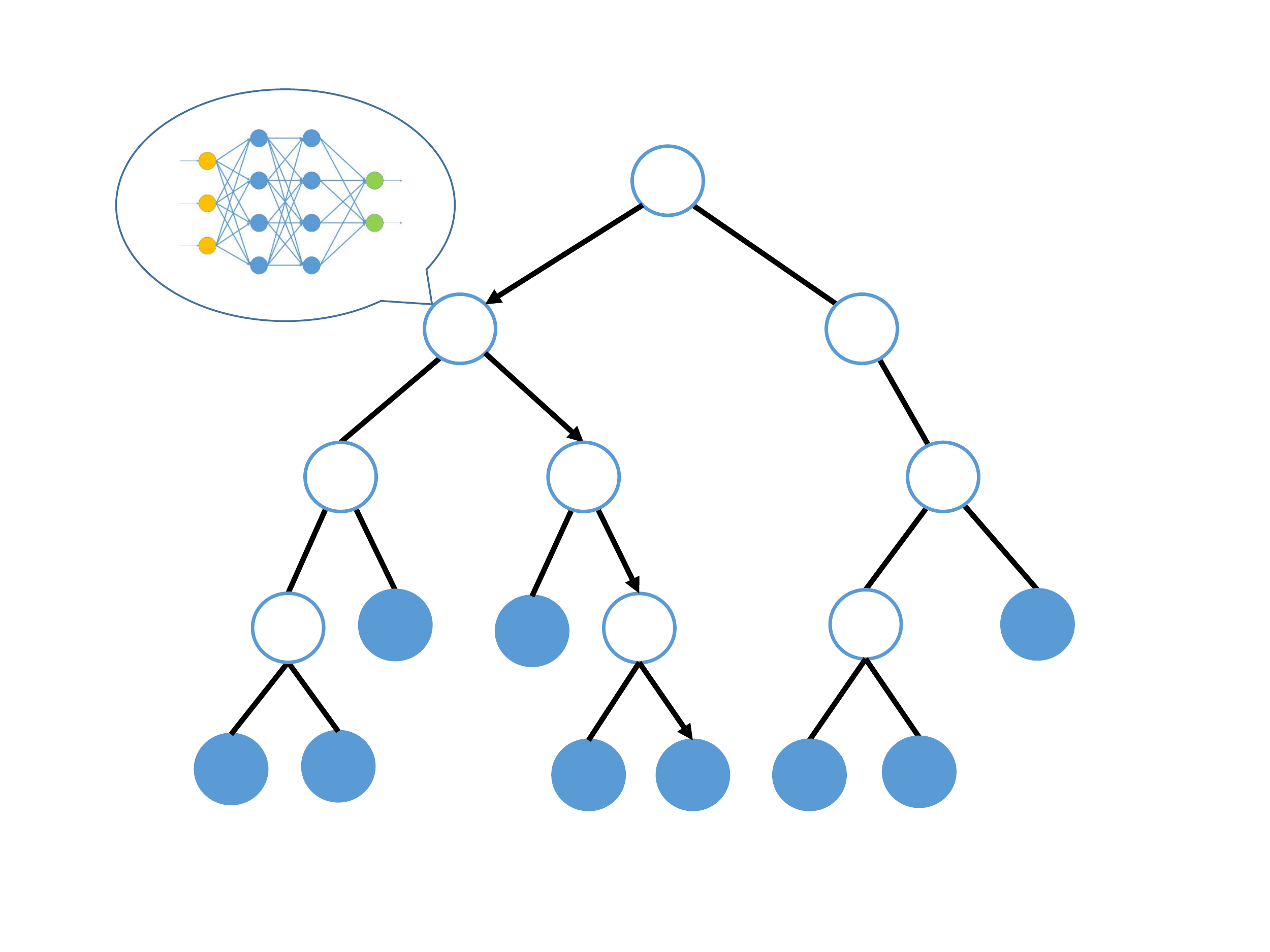}
			\label{fig:opl}
		\end{minipage}%
	}%
	\centering
	\caption{A comparison of the end-to-end learning and optimization policy learning.}
\end{figure}

This investigation sits at the intersection of wireless communications, mathematical optimization, and machine learning. To set the stage, we first review available results on ``learning to optimize'' in these different areas. The goal of ``learning to optimize'' is to obtain near-optimal algorithms with affordable computational complexity for challenging optimization problems. This topic has been developed for decades in the machine learning community, and has also attracted significant attention from mathematical optimization and operation research \cite{yoshua2018machine}.

There are different approaches to apply machine learning methods, and a straightforward one is ``end-to-end learning''. The idea is to regard the optimization algorithm as a black-box, and directly learn the input-output mapping \cite{yoshua2018machine}, as illustrated in Fig. \ref{fig:e2e}. For example, the pointer networks \cite{vinyals2015pointer,bello2016neural} have been proposed for approximating the output of algorithms for combinatorial optimization problems. In wireless networks, the study in \cite{sun2018learning} also applied this idea. The adopted neural networks in these studies have generic architectures and perform well in practical problems. Nevertheless, they require large amounts of samples for training, which impedes using optimal solutions as labels when solving MINLPs. In this approach, the machine learning model is used as an approximator for an existing optimization algorithm, and thus the performance highly depends on the baseline algorithm. Besides, the feasibility of constraints is guaranteed by simply projecting the solution to the feasible set, which will result in performance loss. Furthermore, such simple projection trick will have difficulties in handling general constraints, even convex ones.

One major objective of this study is to develop a ``learning to optimize'' framework with few training problem instances, which falls in the paradigm of \emph{few-shot learning}, a method that mimics the way humans learn new concepts \cite{goodfellow2016deep}. One effective way to reduce the sample complexity is to consider a specific optimization algorithm and exploit its algorithmic structure \cite{yoshua2018machine}. Such ``Optimization policy learning'' approach learns the optimal policy in the given algorithm, and thus has little influence on the feasibility guarantee of that algorithm, as illustrated in Fig. \ref{fig:opl}. It has recently attracted lots of attention, especially for mixed-integer linear programs (MILPs) \cite{he2014learning,khalil2016learning} and problems on graphs \cite{khalil2017learning}. Besides reducing the sample complexity and providing feasibility guarantee, there are other advantages by exploiting the algorithm structure. Firstly, by exploration, the performance of the learned policy is not bounded by the performance of heuristic algorithms that generate labels \cite{yoshua2018machine,songlearning}, which helps to achieve close to the optimal performance. Therefore, we can develop an optimization policy with better performance than the learned one. Secondly, these methods enjoy good generalization abilities, e.g., they can scale up to larger problem sizes and generalize to different problems \cite{he2014learning,khalil2017learning}. Inspired by the unique advantages as mentioned above, our proposed approach is based on ``optimization policy learning'', instead of ``end-to-end learning''. 

Optimization policy learning has been applied to solve combinatorial optimization problems. One such study, which is related to our investigation, is \cite{he2014learning} for MILPs. While \cite{he2014learning} achieves certain speedup compared with the branch-and-bound algorithm for MILPs, its computational complexity is still high and more importantly, it cannot be directly applied to constrained MINLP problems. First, compared with MILPs, it is much more time-consuming to solve the relaxed problems in MINLPs for resource management in wireless networks. Second, in \cite{he2014learning}, to address the feasibility problem, the search space is kept large for each problem in order to guarantee feasibility, which brings additional complexity. In this paper, we have made great efforts to improve the computational efficiency over \cite{he2014learning}. Specifically, to address the first issue, we use neural networks and problem dependent features to avoid solving relaxed problems. To address the second issue, we propose an adaptive strategy to control the search space and guarantee the feasibility, which leads to a small search space in average. These two innovations allow our proposed framework to achieve low computational complexity with near-optimal performance.

\subsection{Contributions}
In this paper, we focus on finding near-optimal solutions for MINLP resource management problems in wireless networks using machine learning. The main innovation is a Learning to Optimize framework for Resource Management, called LORM, which is based on optimization policy learning. The merits of this framework and major contributions of this paper are summarized as follows:
\begin{enumerate}
	\item We develop the LORM framework to solve MINLPs for resource management in wireless networks. The proposed approach accelerates the branch-and-bound algorithm by casting the branch-and-bound algorithm as a sequential decision problem and learning the optimal pruning policy via imitation learning. This framework achieves near-optimal performance with few training problem instances and low computational complexity, by exploiting the structure of the branch-and-bound algorithm and problem data. Furthermore, the feasibility of constraints is guaranteed by iteratively increasing the search space.

	\item To tackle the task mismatch issue, we adopt a transfer learning method via self-imitation in LORM, which leads to the LORM-TL framework. Compared with the traditional transfer learning method, i.e., fine-tuning, self-imitation only requires a few additional \emph{unlabeled} training problem instances. This is achieved by a proposed exploration policy to label the additional training problem instances in the new scenario, followed by fine-tuning the learned policy with these labels.

	\item We test LORM and LORM-TL on a typical MINLP problem, i.e., network power minimization in Cloud-RANs \cite{shi2014group}. Simulations will demonstrate the effectiveness of LORM and LORM-TL in the following aspects:
	\begin{enumerate}
		
		\item LORM outperforms the specialized state-of-the-art methods and achieves near-optimal performance under a variety of system configurations, with only tens of training problem instances. It also achieves significant speedup compared with the branch-and-bound and existing method \cite{he2014learning}.
		
		\item It is demonstrated that, with LORM, a model trained on one setting can achieve state-of-the-art performance in moderately different settings, e.g., with different large-scale fading and different numbers of users or base stations. Such generalization capability is achieved via exploiting the algorithm structure and careful feature design.
		
		\item When the variation of the system configuration becomes dramatic, directly generalizing the model trained via LORM to the new setting induces considerable performance degradation. In this case, LORM-TL is able to achieve comparable performance with the model trained on sufficient samples for the test setting from scratch, and it only relies on tens of additional unlabeled training problem instances.
	\end{enumerate}

\end{enumerate}

\section{Resource Management and Branch-and-Bound}
In this section, we first introduce a general formulation for mixed-integer resource management problems in wireless networks. A global optimization algorithm, i.e., branch-and-bound, is then introduced, followed by some observations.

\subsection{Mixed-integer Resource Management}
A wide range of resource management problems in wireless networks can be formulated as MINLP problems, which consist of a discrete optimization variable $\vct{a}$, e.g., indicating cell association or subcarrier management, and a continuous optimization variable $\vct{w}$, e.g., transmit power, subject to resource or performance constraints. Typical examples include subcarrier management in OFDMA \cite{wong1999multiuser}, user association \cite{ye2013user}, network power minimization in Cloud-RANs \cite{shi2014group}, and joint task offloading scheduling and transmit power management \cite{mao2016dynamic}. A general formulation for such problems is given by
\begin{equation}
\begin{aligned}
\mathscr{P}: &\underset{\vct{a},\vct{w}}{\text{minimize}}
& & f(\vct{a},\vct{w})\\
& \text{subject to}
& & \mathcal{Q}(\vct{a},\vct{w})\leq 0\\
&
& & a[i] \in \mathbb{N}, w[i] \in \mathbb{C}, 
\end{aligned}
\end{equation}
where $f(\cdot,\cdot)$ is an objective function, e.g., the sum rate or network power consumption, $a[i]$ and $w[i]$ are the elements of $\vct{a}$ and $\vct{w}$ respectively, and $\mathcal{Q}(\cdot,\cdot)$ represents constraints such as the QoS or power constraint. MINLP problems are NP-hard in general.

\subsection{Branch-and-Bound}\label{sec:BnB}

\begin{figure*}[htb]
	\centering
	\includegraphics[width=0.7\textwidth]{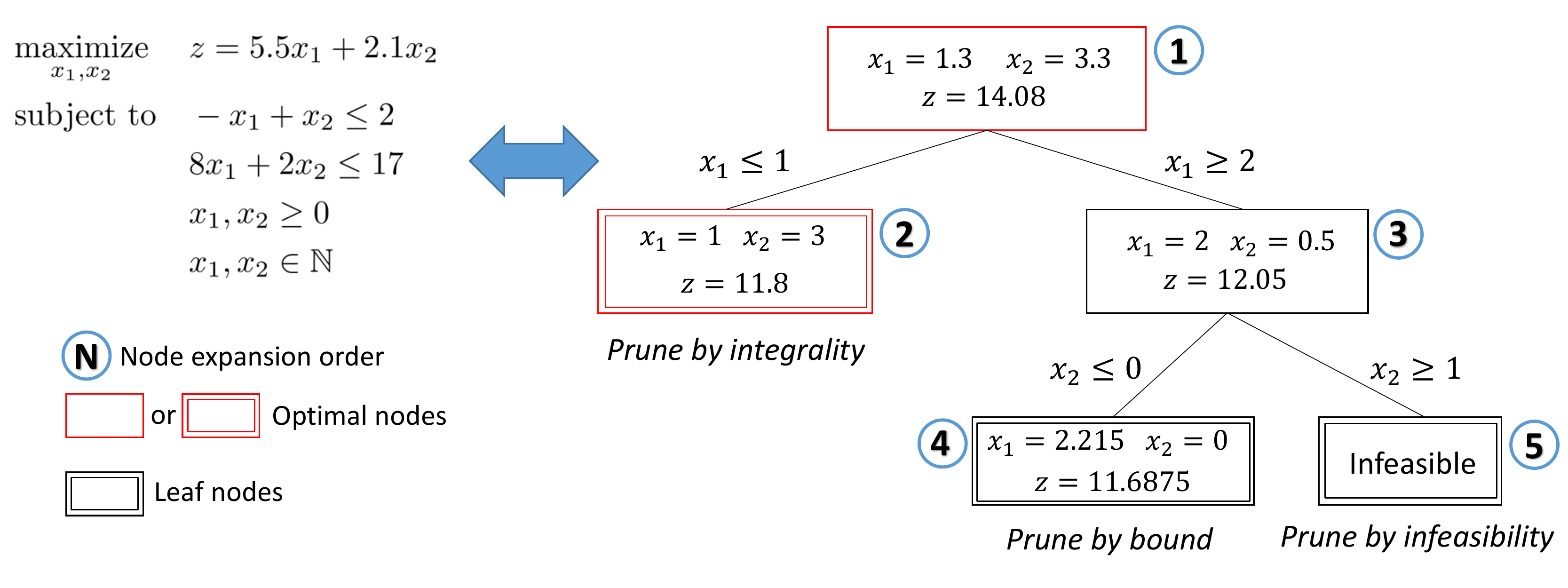}
	\caption{An example of the branch-and-bound algorithm \cite{Conforti2014integer}. The solution of the relaxed linear programming problem is $x_1=1.3$, and  $x_2=3.3$. It is used as the initial solution for the tree search. The objective value $z=14.08$ provides an upper bound to the original problem. According to the branching rule, we branch out to the first variable $x_1$ and create two sub-problems. According to the node selection policy, we select the left child. The solution to the relaxed problem of the left child is $x_1=1$ $x_2=3$. Because they are integer values, we prune the node by integrality. The objective value $z=11.8$ provides a lower bound for the original problem. We then turn to the right child of the root node. We solve it and branch out on $x_2$, which generates two child nodes. The optimal objective value of the left child is $11.6875<11.8$, so we prune it by bound. The relaxed problem of the right child is infeasible and hence we prune it by infeasibility. Thus, the whole branch-and-bound tree is visited and we get the solution to the original problem as $x_1=1$ $x_2=3$.}
	\label{fig:BnB}
\end{figure*}

\begin{algorithm}  
	\caption{The branch-and-bound algorithm.}  
	\label{alg:BnB}  
	\small
	\begin{algorithmic}[1]
		\State $\mathcal{L} \leftarrow \{{\rm {N}}_1\},c^* \leftarrow + \infty,t \leftarrow 0$
		\While{$\mathcal{L} \neq \emptyset$}
		\State $t \leftarrow t+1$
		\State ${\rm {N}}_t \leftarrow$ selects a node from $\mathcal{L}$
		\State $\mathcal{L} \leftarrow \mathcal{L}\backslash\{{\rm {N}}_t\}$
		\State $(\vct{a}^*_t,\vct{w}^*_t),z^*_t \leftarrow$ solve the relaxed problem of $\mathscr{P}_t$
		\If{$\pi_p({\rm {N}}_t,\vct{a}^*_t,z^*_t)$ = \emph{preserved} }
		\State ${\rm {N}}_t^1,{\rm {N}}_t^2 \leftarrow$ two children of ${\rm {N}}_t$
		\State $\mathcal{L} \leftarrow \mathcal{L} \cup \{{\rm {N}}_t^1,{\rm {N}}_t^2\}$
		\EndIf
		\If{$a^*_t[j],\forall j$ is integer}
		\State $c^* \leftarrow \min(c^*,z^*_t)$
		\EndIf
		\EndWhile
	\end{algorithmic}  
\end{algorithm} 
The branch-and-bound algorithm is one of the state-of-the-art algorithms to find an optimal solution to the MINLP problems \cite{Conforti2014integer}. The discrete variables and the continuous variables are first separated. The algorithm then determines the values of the integer variables via a tree search process, followed by obtaining the values of the continuous variables via a convex solver. The tree search process consists of three main policies: Node selection policy, variable selection policy, and pruning policy, which together iteratively construct a binary search tree. 

We first specify some notations. Each node in the binary search tree contains a MINLP problem and its relaxed problem. Let ${\rm {N}}_t$ denote the node selected at the $t$-th iteration, $\mathscr{P}_{t}$ denote the MINLP problem at node ${\rm {N}}_t$, and $\mathcal{G}_{t}$ represent the feasible set of $\mathscr{P}_{t}$. The natural relaxation of  $\mathcal{G}_{t}$, i.e., relax all the integer constraints into box constraints, is denoted as $\mathcal{H}_{t}$. For example, binary constraints $\mathcal{G}_{t} = \{(\vct{a},\vct{w}): a[i] \in \{0,1\}, \forall i \}$ can be relaxed into box constraints $\mathcal{H}_{t} = \{(\vct{a},\vct{w}): a[i] \in [0,1], \forall i \}$. Thus, the original problem and relaxed problems at node ${\rm {N}}_t$ can be represented respectively as $\underset{\vct{a},\vct{w}}{\text{minimize}}\{f(\vct{a},\vct{w}):(\vct{a},\vct{w}) \in \mathcal{G}_{t}  \}$ and $\underset{\vct{a},\vct{w}}{\text{minimize}}\{f(\vct{a},\vct{w}):(\vct{a},\vct{w}) \in \mathcal{H}_{t}  \}$. Let $(\vct{a}^*_t,\vct{w}^*_t)$ and $z^*_t$, respectively, denote the optimal solution and the optimal objective value of the relaxed problem at ${\rm {N}}_t$, i.e., 
\begin{equation}
\begin{aligned}
	&(\vct{a}^*_t,\vct{w}^*_t) = \underset{\vct{a},\vct{w}}{\argmin}\{f(\vct{a},\vct{w}):(\vct{a},\vct{w}) \in \mathcal{H}_{t}  \} \\
	&z^*_t = \underset{\vct{a},\vct{w}}{\min}\{f(\vct{a},\vct{w}):(\vct{a},\vct{w}) \in \mathcal{H}_{t}  \}. 
\end{aligned}
\end{equation}
We use $a^*_t[j]$ to represent the $j$-th element of $\vct{a}^*_t$. The best integer solution to $\mathscr{P}$ found before or in this iteration is denoted as $c^*$. 

The branch-and-bound algorithm maintains an unexplored node list $\mathcal{L}$, and this list only contains the root node ${\rm {N}}_1$ at the beginning. The MINLP problem at the root node is $\mathscr{P}$. At each iteration, the node selection policy first selects a node and pops it from the unexplored list. The relaxed problem at this node is solved to obtain $\vct{a}^*_t$ and $z^*_t$. The pruning policy $\pi_p$ then determines whether to preserve this node according to $\vct{a}^*_t$ and $z^*_t$. The algorithm enters the next iteration if the pruning policy decides to prune this node, i.e., no child node of this node will be considered. Otherwise, the variable selection policy chooses a variable index $j$ such that $a^*_t[j]$ is fractional. Then, the feasible set $\mathcal{G}_{t}$ is divided into two parts as
\begin{equation}
\begin{aligned}
	\mathcal{G}^{1}_{t} = \mathcal{G}_{t} \cap \{(\vct{a},\vct{w}) :a[j] \leq \lfloor a^*_t[j] \rfloor \}, \\
	\mathcal{G}^{2}_{t} = \mathcal{G}_{t} \cap \{(\vct{a},\vct{w}) : a[j] \geq \lceil a^*_t[j] \rceil \}.
\end{aligned}
\end{equation}
Two new MINLP problems are formed based on the partition
\begin{equation}
\begin{aligned}
\mathscr{P}^{1}_{t} : \underset{\vct{a},\vct{w}}{\text{minimize}}\{f(\vct{a},\vct{w}):(\vct{a},\vct{w}) \in \mathcal{G}_{t}^{1}  \}, \\
\mathscr{P}^{2}_{t} : \underset{\vct{a},\vct{w}}{\text{minimize}}\{f(\vct{a},\vct{w}):(\vct{a},\vct{w}) \in \mathcal{G}_{t}^2  \}. 
\end{aligned}
\end{equation}
Two child nodes ${\rm {N}}_t^1$ and ${\rm {N}}_t^2$, whose MINLP problems are $\mathscr{P}^{1}_{t}$ and $\mathscr{P}^{2}_{t}$, respectively, are produced and put into the unexplored node list $\mathcal{L}$. These procedures repeat at each iteration until there are no nodes in $\mathcal{L}$. The pseudo-code is shown in Algorithm \ref{alg:BnB} and an illustration in shown in Fig. \ref{fig:BnB}.

In the standard branch-and-bound algorithm, the pruning policy $\pi_p$ prunes a node only if one of the following conditions is met:

\begin{itemize}
	\item \emph{Prune by bound}: $z^*_t \geq c^*$. $z^*_t$ provides a lower bound for the optimal objective value of $\mathscr{P}_t$ since $\mathcal{G}_t \subseteq \mathcal{H}_t$. Under such circumstance, the optimal solution to $\mathscr{P}_t$ cannot be the best solution as the lower bound is worse than the best objective value found.
	\item \emph{Prune by infeasibility}: The relaxed problem of $\mathscr{P}_t$ is infeasible. This is a special case of \emph{prune by bound} if the infeasibility is viewed as $z^*_t = +\infty$.
	\item \emph{Prune by integrality}: $a^*_t[j],\forall j,$ is an integer. It is unnecessary to search the children of ${\rm {N}}_t$ since we have found the optimal solution to $\mathscr{P}_t$.
\end{itemize}

\subsection{Observations}\label{sec:prune}
Branch-and-bound has been widely employed in solving MINLPs in wireless networks, as it is capable to obtain globally optimal solutions. However, its computational complexity is exponential in the dimension of $\vct{a}$, which cannot be tolerated in practice.

The heavy computation of the branch-and-bound algorithm comes from two aspects. 
\begin{enumerate}
	\item The number of nodes in the branch-and-bound tree is exponential in the dimension of $\vct{a}$;
	\item Solving relaxed problems in MINLPs, e.g., second order cone programs (SOCPs) or semidefinite programs (SDPs), is computationally expensive.
\end{enumerate}

These two observations form the basis for our proposed method. As the number of nodes in the search tree is controlled by the pruning policy, improving the policy helps to reduce the computation complexity. The pruning policy in the standard branch-and-bound algorithm is conservative because the optimality must be guaranteed, which is achieved by checking that all the other solutions are worse than the returned one. Therefore, most of the running time of the branch-and-bound algorithm is spent on checking the non-optimal nodes. Thus, we can dramatically boost the efficiency of the branch-and-bound algorithm by learning an aggressive pruning policy, without enforcing the optimality guarantee. 

For the second issue, in the standard branch-and-bound algorithm, the relaxed problem must be solved because the features of the pruning policy merely consist of the solution of the relaxed problem. We can reduce the number of relaxed problems if we train the classifier based on the problem data and only the solution of the relaxed problem at the root node. While the tree size reduction via aggressive pruning has been discussed and attempted in previous studies, e.g., \cite{he2014learning}, our proposal to reduce the relaxed problems is new, and it can greatly speed up the computation.

\section{LORM: Learning to Optimize for Resource Management}
In this section, we first introduce the idea of learning the policies in the branch-and-bound algorithm via imitation learning. Then key components of the LORM framework, including classifier design, feature design, and algorithm design are then presented, followed by a complexity analysis and some discussions.

\subsection{Learning Policies via Imitation Learning}\label{sec:why} \label{sec:fw}
The search procedure of the branch-and-bound algorithm can be regarded as a sequential decision problem. In theory, the policy in such finite-dimensional sequential decision problems can be learned via supervised learning if we have features and labels for all the nodes. However, this direct approach is inefficient because there is an exponential number of nodes, which makes it tedious to obtain all the features. A too large dataset will also increase the complexity of training. To reduce the training effort and retain good performance, we instead employ DAgger (Data Aggregation) \cite{ross2011reduction} to collect features and labels dynamically. This leads to \emph{imitation learning}, a machine learning method that enjoys low sample complexity and good generalization capability \cite{osa2018algorithmic}. In the following, key steps to apply imitation learning to learn the pruning policy will be introduced. The design of the classifier and features, as well as other implementation issues, will be discussed in the following subsections. One can refer to \cite{osa2018algorithmic} for a more detailed background on imitation learning.

Modeling the tree search process as a sequential decision problem is a common method in machine learning \cite{silver2016mastering}. A sequential decision problem consists of a state space $\mathcal{X}$, an action space $\mathcal{Q}$, and a policy space $\Pi$. A single trajectory $\tau$ consists of a sequence of states $x_1, \cdots, x_T \in \mathcal{X}$, a sequence of actions $q_1$,$\cdots$,$q_T \in \mathcal{Q}$, and a policy $\kappa \in \Pi$ and $\kappa(\cdot):\mathcal{X} \rightarrow \mathcal{Q}, x_t \mapsto q_t$. The agent uses its own policy $\kappa$ to make a sequence of decisions $q_t, t = 1,\cdots,T,$ based on the state it observed, i.e., $x_t,i=1,\cdots,T$. Specifically, at time step $t$, the agent is aware of the state $x_t$. It then uses its policy to take an action $\kappa(x_t)=q_t$. The action $q_t$ might have some influences on the environment, i.e., the next state $x_{t+1}$ is influenced by $q_t$. 

The procedure of the branch-and-bound algorithm, essentially a tree search process, is a sequential decision problem because the pruning policy should make a decision at each iteration sequentially. We define $\phi$ as the feature map, i.e., $\phi(\cdot):\mathcal{X} \rightarrow \mathbb{R}^*$. When learning the pruning policy, i.e., to learn whether to prune a node ${\rm {N}}_t$, the state $x_t$ is the feature of ${\rm {N}}_t$, denoted as $\phi({\rm {N}}_t)$, and the action is a binary value, i.e., $q_t \in \{prune, preserve\}$. Thus, the policy $\kappa$ is a binary classifier that takes $\phi({\rm {N}}_t)$ as the input and outputs a binary value. We hope the learned policy can find the optimal solution with the minimal number of nodes expanded. Therefore, only the nodes whose feasible regions containing the optimal solutions should be preserved. These nodes are referred to as \emph{optimal nodes}. We then label all the optimal nodes as $preserve$ and other nodes as $prune$.

With the feature and label of every node, the policy can be learned via supervised learning. However, directly applying supervised learning is impractical even for medium-size wireless networks, as there would be a huge number of nodes. Meanwhile, including only part of the nodes will degrade the performance. Imitation learning differs from supervised learning in the way of collecting features and labels for the classifier. For supervised learning, all the features and labels are needed before the training process. In contrast, imitation learning iteratively collects data to correct the mistakes the learned policy makes, and the data collection algorithm is called DAgger \cite{ross2011reduction}. With DAgger, at the $i$-th iteration, the standard policy in the branch-and-bound algorithm is replaced with the learned policy $\pi^{(i)} \in \Pi$. We collect the features of all the nodes explored by the learned policy and their corresponding labels. Then a new policy $\pi^{(i+1)}$ is learned via supervised learning on the dataset collected at previous iterations, and it corrects the mistakes made by $\pi^{(i)}$. In this way, the learned policy approaches the optimal policy progressively. The detailed algorithm for DAgger can be found as Algorithm 3.1 in \cite{ross2011reduction}.

\subsection{Neural Networks as Classifiers}\label{sec:mlp}
In this paper, adopting imitation learning, we propose to use a neural network as the classifier for the pruning policy. One advantage of neural networks is that we can tune a parameter to control the computation-performance trade-off. This will not only enable us to use an iterative algorithm to guarantee the feasibility and outperform non-optimal labels, but also assist the development of the transfer learning approach in Section IV.

Recall that learning the pruning policy is a binary classification problem, where the input is the feature vector of the node and the output is a binary class in $\{prune,preserve\}$. We employ an $L$-layer multi-layer perceptron (MLP), a type of neural networks, as the binary classifier with $L$-layer. The $k$-th layer's output is calculated as:
\begin{equation}
\vct{g}^k = \text{Relu}(\mtx{W}^k\vct{g}^{k-1}+\vct{b}^k),
\end{equation}
where $\mtx{W}^k$ and $\vct{b}^k$ are the learned parameters of the $k$-th layer. $\vct{g}^k,k=1,\cdots,L,$ denotes the output of the $k$-th layer, and $\vct{g}^0$ is the input feature vector, which will be designed in the next subsection. $\text{Relu}(\cdot)$ is the rectified linear unit function, i.e., $\max(0,\cdot)$. The output indicates the probability of each class, which is a normalization of the last layer's output $\vct{g}^L \in \mathbb{R}^2$:
\begin{equation}
\vct{e}[i] = \frac{\exp(\vct{g}^L[i])}{\sum_{j=1,2}\exp(\vct{g}^L[j])},i=1,2,
\end{equation}
where $\vct{e}[i]$ indicates the $i$-th component of vector $\vct{e}$.

In the training stage, the loss function is the weighted cross entropy, given by:
\begin{equation}
\ell = -\sum_{j=1,2} \vct{w}[j] \cdot \vct{y}[j]\log(\vct{e}[j]),
\end{equation}
where $\vct{y}$ is the label, i.e., $\vct{y}=(1,0)$ indicates that it belongs to the class \emph{prune}, and $\vct{y}=(0,1)$ otherwise. $\vct{w}$ denotes the weight of each class, which is tuned by hand. Two parts contribute to the weight. First, if the number of non-optimal nodes is much larger than the number of optimal nodes, we should assign a higher weight to the class \emph{preserve} in order to let the neural network not to ignore this class. We denote this part as $\vct{o}_1$ and it can be computed by:
\begin{equation}
\begin{aligned}
\vct{o}_1[1] &= \frac{\text{\# optimal nodes in the training set}}{\text{\# nodes in the training set}}, \\ 
\vct{o}_1[2] &= 1 - \vct{o}_1[1].
\end{aligned}
\end{equation}
Second, when the number of feasible solutions is small, we should assign a higher weight to the optimal nodes in the training dataset in order not to miss good solutions. This parameter, denoted as $o_2$, is tuned by hand to achieve good performance on the validation dataset. The total weight is calculated by $\vct{w} = \vct{o}_1 \odot \vct{o}_2$, where $\odot$ is a hadamard product.

The classifier controls whether to prune a node. Pruning too aggressively leads to finding no feasible solutions. Pruning too moderately will lead to a large number of preserved nodes and high computation cost. Thus, a desirable property of the classifier is an ability to dynamically control the search space during the test stage.

\paragraph*{Control the search space} During the test phase, the MLP outputs $\vct{e}$, which indicates the probability of each class, followed by setting a threshold indicating which class it belongs to. The threshold for the class \emph{prune} is denoted as $\Lambda$. In the standard classification problem, $\Lambda = 0.5$. If e[1] > $0.5$, the input should belong to the first class. Otherwise, it should belong to the second class. We can adjust the threshold to control the search space. Specifically, a pruning policy with a higher threshold will preserve more nodes than that with a lower threshold, which leads to a larger search space and better performance. In the simulations, we increase the threshold iteratively if we can not get any feasible solution. Specifically, we increase the threshold in a moderately exponential way, i.e., $\Lambda_k = 1 - 0.5 \cdot 0.8^k$ at the $k$-th iteration. The specific method is shown in Algorithm \ref{alg:new_BnB}.
This method is shown to be very effective in the simulations, and the total number of iterations is smaller than $3$ when the training task and test task are the same.

\subsection{Feature Design}\label{sec:fea}
The input features of the neural network play a critical role in its accuracy, which determines the performance and computational complexity of the proposed framework. As a result, they should be carefully designed by leveraging domain knowledge. The available information consists of the problem data and the structure of the binary search tree. We propose to include two types of features, problem-independent features and problem-dependent features, which correspond to the structure of the tree and problem data respectively.

\subsubsection{Problem-independent Features}
These kinds of feature only contain information about the branch-and-bound tree and the solutions and objective values of the relaxed problems. They are universal to all MINLP problems and are invariant from problem to problem. Thus, the learning framework will have a good generalization capability if we only use problem-independent features. Such features can be categorized into three groups:
\begin{enumerate}
	\item Node features, computed merely from the current node ${\rm {N}}_i$. They contain the depth of ${\rm {N}}_i$ and the optimal objective value of the relaxed problem $z_i$.
	
	\item Branching features, computed from the branching variable. They contain the branching variable's value at the current node, i.e., $a^*_i[j]$, that in the relaxed problem at its father node, and that in the relaxed problem of the root node, i.e., $a^*_1[j]$.
	
	\item Tree features, computed from the branch-and-bound search tree. They contain the optimal objective value at the root node, i.e., $z_1$, the number of solutions found ever, and the best objective value found ever, i.e., $c^*$.
\end{enumerate}

Due to the significant variations among the objective value of $\mathscr{P}$ under different network settings, all the objective values used as features in the branch-and-bound search tree should be normalized by the optimal objective value of the relaxed problem at the root node.

\subsubsection{Problem-dependent Features} Problem-independent features are universal to all MINLP problems. Unfortunately, we cannot learn efficient policies with only problem-independent features, as proved in \cite{Balcan18branch}. Thus, the problem-dependent features, are introduced. In a general MINLP, the problem-dependent features can be the coefficients of integer variables. As we consider problems in wireless communications, domain knowledge can be exploited and the problem dependent features can be channel state information (CSI) or some descriptions of the radio resources, e.g., power features. The design of problem dependent features for the network power minimization in Cloud-RANs is described in Section \ref{sec:simu}. To maintain a good generalization capability, problem-dependent features are also normalized.

\subsection{The LORM Framework} \label{sec:imp}
In this subsection, we present the algorithm of the LORM framework in details, which consists of a training phase and a test phase.

\paragraph{The training phase} The DAgger algorithm is used to collect data and train the classifier during the training phase. DAgger is an iterative algorithm, with a total of $M$ iterations. At the $i$-th iteration, we have a learned policy $\pi^{(i-1)}$ (we set $\pi^{(0)}$ as a random policy). We use a modified branch-and-bound algorithm to collect data and train the classifier by using the depth-first node selection policy, depth-first variable selection policy, and the learned pruning policy $\pi^{(i-1)}$. For each problem instance, the relaxed problem at the root node is firstly solved to get the problem-independent features $(\vct{a}^*_1,\vct{w}^*_1),z^*_1$. A variable is selected, and two children nodes of the root node are expanded. For each child ${\rm {N}}_t$, $\pi^{(i-1)}$ is used to decide whether to prune or preserve the node based on the problem-dependent feature and $(\vct{a}^*_1,\vct{w}^*_1),z^*_1$. We then add the node feature and corresponding labels of the node $(\phi({\rm {N}}_t),\kappa^*({\rm {N}}_t))$ into a set $\mathcal{D}$. After all the nodes in this problem instance are explored or pruned, the optimal nodes are also added into $\mathcal{D}$. This is because the optimal nodes represent the desired behavior and we do not want to miss any of them. The new policy $\pi^{(i)}$ is trained on $\mathcal{D}$ with supervised learning after all the problem instances are solved via the modified branch-and-bound algorithm. Thus, a sequence of policies $\pi^{(0)},\cdots,\pi^{(M)}$ are obtained and we select the policy performs best on the validation dataset in the test. 
\begin{algorithm}  
	\small
	\caption{LORM Branch-and-bound($\kappa$)}  
	\label{alg:new_BnB} 
	\begin{algorithmic}[1]
		\State $\mathcal{L}\leftarrow \{{\rm {N}}_1\},\mathcal{D}\ \leftarrow \{\},t \leftarrow 0, c^* = + \infty$
		\State $(\vct{a}^*_1,\vct{w}^*_1),z^*_1 \leftarrow$ solve the relaxed problem of $\mathscr{P}_{1}$
		\While{$c^* = \infty$}
		\State Increase the threshold $\Lambda$ of $\kappa$
		\While{$\mathcal{L}\neq \emptyset$}
		\State $t \leftarrow t+1$
		\State ${\rm {N}}_t \leftarrow$ select a node from $\mathcal{L}$
		\If {${\rm {N}}_t$ is leaf node}
		\State $(\vct{a}^*_t,\vct{w}^*_t),z^*_t \leftarrow$ solve and save the relaxed problem of $\mathscr{P}_t$ or load from $\mathcal{T}$
		\State $c^* \leftarrow \min(c^*,z^*_t)$
		\Else
		\State $f \leftarrow \phi({\rm {N}}_t)$
		\If {$\kappa(f) = preserve$}
		\State ${\rm {N}}_t^1,{\rm {N}}_t^2 \leftarrow$ two children of ${\rm {N}}_t$
		\State $\mathcal{L} \leftarrow \mathcal{L} \cup \{{\rm {N}}_t^1,{\rm {N}}_t^2\}$
		\EndIf
		\EndIf
		\EndWhile
		\EndWhile
		\State \Return $c^*$
	\end{algorithmic}  
\end{algorithm}
\paragraph{The test phase} During the test phase, we replace the standard pruning policy in the branch-and-bound algorithm with the learned policy $\kappa$. For a non-leaf node ${\rm {N}}_t$, the learned policy $\kappa$ determines whether to prune the node or not. For the leaf node, the subproblem becomes a convex problem since all the integer variables are determined. It is solved via some convex solvers to check the feasibility and obtain the values for continuous variables. However, this may produce infeasible solutions if the pruning policy is too aggressive. Recall that in Section \ref{sec:mlp}, we propose to tune a parameter to control the search space. Specifically, we can iteratively increase the threshold $\Lambda$ if the problem is infeasible. During this process, the convex problem at the leaf node needs to be solved several times, which is not time-efficient. We propose to build a lookup table to accelerate this process. The lookup table is indexed by the value of integer variables and is used to store the solutions of the relaxed problems. If we encounter a relaxed problem that has not been solved before, we solve it and save this problem instance and its solution into the lookup table. Otherwise, we directly extract the solution from the lookup table. The pseudo-code for our proposed algorithm is shown in Algorithm \ref{alg:new_BnB}.

\subsection{Complexity Analysis}\label{sec:ana}
In this subsection, we present some complexity analysis for the LORM framework. We follow the common assumptions and analysis for imitation learning \cite{ross2011reduction}, and study the expected number of nodes explored, which indicates the number of samples we have, and the expected number of convex problems solved, which indicates the computational complexity of the proposed algorithm.

\begin{thm}\label{thm:socp}
	Given the number of integer variables $L$, a node pruning policy which expands a non-optimal node with probability $\epsilon_1$ and prunes an optimal node with probability $\epsilon_2$, the expected number of nodes explored and the number of relaxed problems solved are $\mathcal{O}(L^2)$ and $\mathcal{O}(L)$, respectively, when $\epsilon_1 \leq 0.5, \epsilon_2 \leq 1$. The expected number of nodes explored and the number of relaxed problems solved are $\mathcal{O}(L)$ and $\mathcal{O}(1)$, respectively, when $\epsilon_1 \leq 0.3, \epsilon_2 \leq 1$.

\end{thm}

As shown in Theorem \ref{thm:socp}, the number of convex problems to solve is $\mathcal{O}(L)$ if we have a proper choice of parameters and features. Thus, LORM enjoys a much lower computational complexity compared with the standard branch-and-bound algorithm. When LORM is applied to the network power minimization problem in Cloud-RANs, which is mixed integer second order cone programming, the expected time complexity is $\mathcal{O}(L^4K^3 + L^3K^2\log\frac{1}{\epsilon})$ if the ADMM-based convex optimization solver \cite{shi2015large} is used. This is much more efficient than the standard branch-and-bound, and comparable to the state-of-the-art method for this problem \cite{shi2014group}, but with better performance, as to be shown in the simulations.

We further make a computational complexity comparison with the method in \cite{he2014learning}, referred to as ``MILP-SVM'', by comparing the number of relaxed problems to solve. The complexity analysis of their method is shown in Theorem I in \cite{he2014learning}. In addition to $\epsilon_1$ and $\epsilon_2$, they assume that the node selection policy ranks some non-optimal node higher than an optimal node with probability $\epsilon$. The number of relaxed problems in MILP-SVM to solve then goes to $\infty$ compared to $1 + \epsilon_1 L$ in LORM when $\epsilon_1$ goes to $0.5$. The number of relaxed problems in MILP-SVM to solve is $\mathcal{O}(L)$ compared to $\mathcal{O}(1)$ in LORM when $\epsilon_1\leq \frac{1}{3}$. This shows that LORM achieves significant speedup compared with MILP-SVM.

The number of training samples for the classifier is the sum of nodes explored in the branch-and-bound tree of each problem instance. Note that the number of problem instances is $|\mathcal{P}|$ and the number of nodes is $\mathcal{O}(L^2)$. Therefore, the number of training samples for the classifier is $\mathcal{O}(L^2|\mathcal{P}|)$.

\subsection{Discussions on LORM}
The LORM framework enjoys a few unique advantages. Specifically, 1) it has the capability to generalize to different settings; 2) it is able to learn the optimization policy with few problem instances; and 3) can work with different baseline algorithms, as explained below.

\paragraph*{Generalization to different settings} As the input and output dimensions of most learning algorithms, e.g., MLP, in the test stage must be the same as in the training stage, most of the end-to-end learning methods \cite{sun2018learning,liang2018towards,lee2018deep} cannot generalize to settings with different problem sizes than those in the training dataset. On the other hand, the input and output dimensions of the policies in the branch-and-bound algorithm are invariant when the problem size changes. Thus, generalization to different problem sizes is not a problem when learning the policies in the branch-and-bound algorithm. Such generalization capability also relies on the design of features (as shown in Section \ref{sec:fea}), and will be tested via simulations.

\paragraph*{Few-shot learning via learning policies} LORM works with few problem instances for two reasons. The first reason is that the available number of samples for learning the classifier is much larger than the problem instances. According to the analysis in Section \ref{sec:ana}, the number of samples is $\mathcal{O}(L^2|\mathcal{P}|)$, where $L$ is the number of integer variables and $|\mathcal{P}|$ is the number of problem instances. Thus, solving one problem instance produces many labeled samples for policy learning. The second reason is that to learn the pruning policy is much easier than directly learning the input-output mapping of an optimization algorithm, which is typically with high dimensional inputs and outputs. These two reasons allow us to use a much lighter-weighted learning algorithm, with which we can achieve few-shot learning.

\paragraph*{Sub-optimal labels} Since the branch-and-bound is time-consuming, we may also generate labels by sub-optimal algorithms. Suppose the solution obtained by the sub-optimal algorithm is $\vct{a}^*, \vct{w}^*$, and we label all the nodes whose feasible sets containing $\vct{a}^*, \vct{w}^*$ as \emph{preserve} and other nodes as \emph{prune}. As the tree search procedure is preserved, the LORM framework still applies, and thus it can work with different available algorithms for a given MINLP problem.

\section{LORM-TL: Transfer Learning via Self-imitation}
While the LORM framework helps to achieve few-shot learning for MINLP resource management problems, its practical implementation still faces a few key challenges, which will be illustrated in this section. We will propose a transfer learning method to address these challenges.
\subsection{Limitations of Machine Learning Methods} \label{sec:ft}
\begin{figure}[htbp]
	\centering
	\subfigure[The key difference between transfer learning and traditional machine learning is that transfer learning can tackle the task mismatch problem with few additional training samples. This is achieved by transferring the knowledge form the old task into the new task.]{
		\begin{minipage}[t]{0.47\linewidth}
			\includegraphics[width=0.7\linewidth]{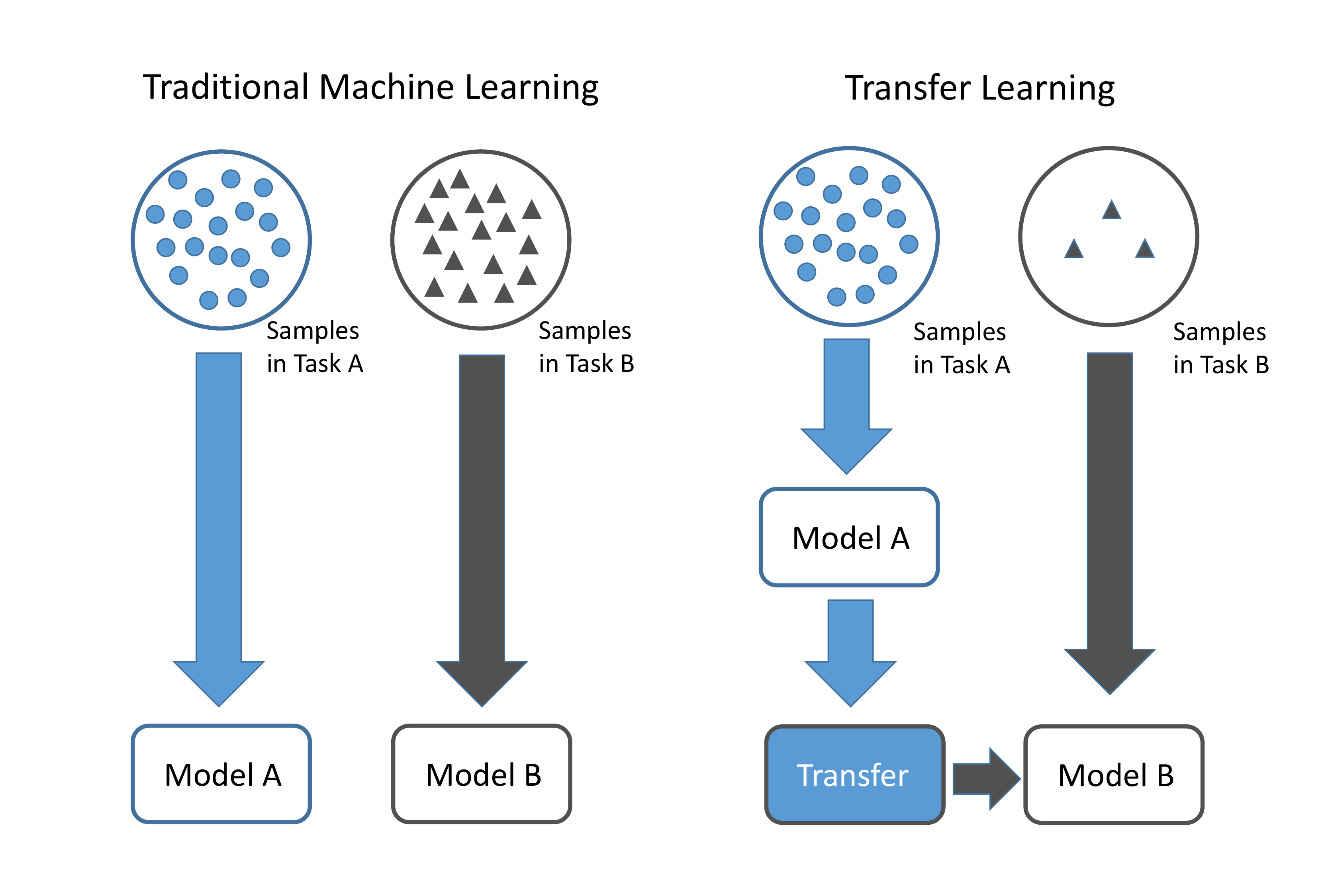}
			\label{fig:tl} 
		\end{minipage}%
	}%
	\hspace{.1in}
	\subfigure[An illustration of transfer learning via fine-tuning. The neural network for the new task keeps the first several layers from the neural network for the old task. The last several layers are trained from scratch or tuned from those for the old task with a small learning rate.]{
		\begin{minipage}[t]{0.47\linewidth}
			\centering
			\includegraphics[width=0.7\linewidth]{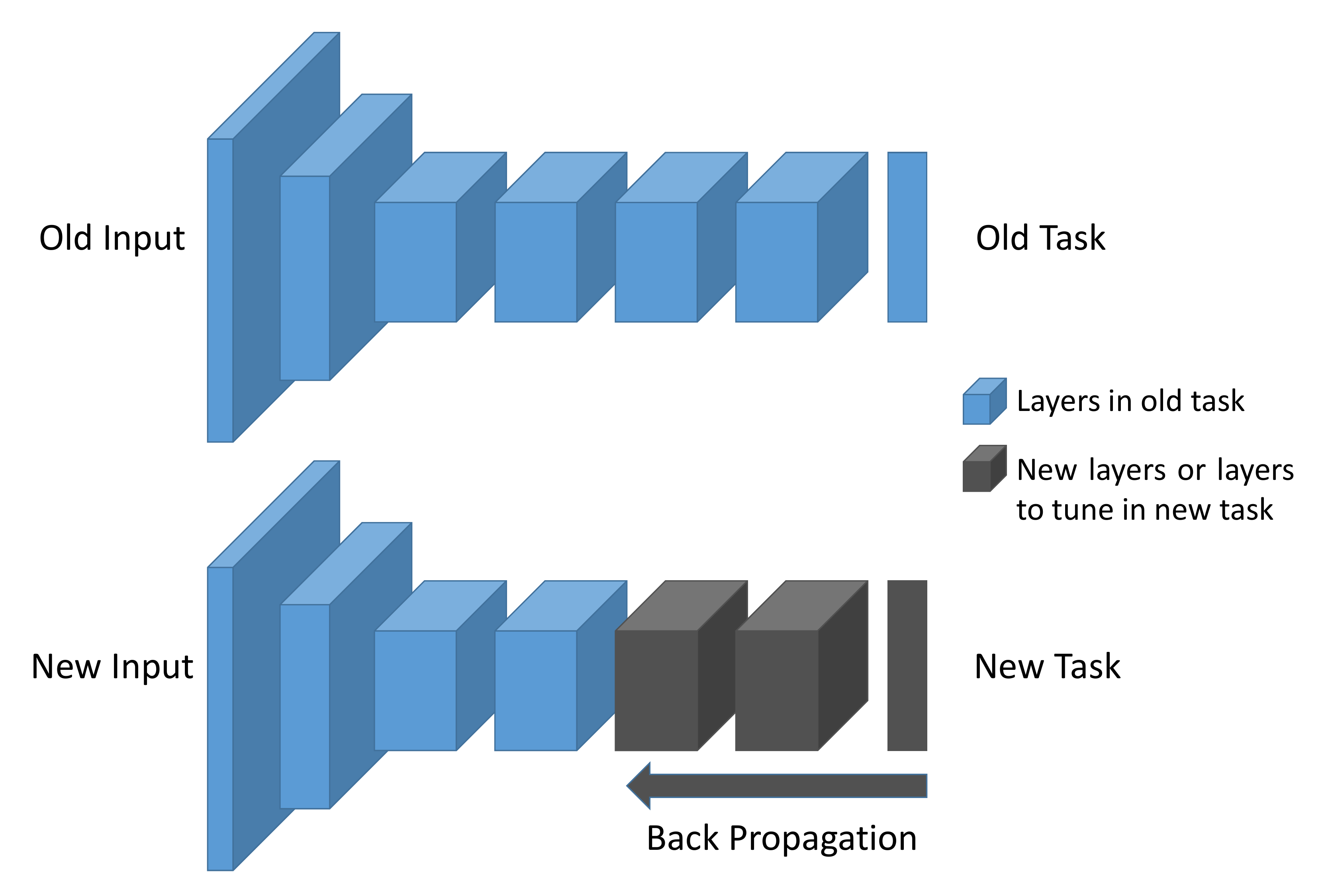}
			\label{fig:ft}
		\end{minipage}%
	}%
	\caption{Illustrations of transfer learning.}
\end{figure}

An essential assumption of most machine learning algorithms is that the training and future testing data are in the same feature space with the same distribution, i.e., the same task \cite{pan2010survey}.  Narrowing down to problems in wireless networks, the ``distribution'' includes ``structure'', e.g., large-scale fading and signal-to-noise-ratio (SNR), and ``size'', e.g., the numbers of users, base stations, and antennas. The performance deteriorates when the machine learning task to be tested is different from the trained one. This is referred to as the \emph{task mismatch} problem. How much will the performance deteriorate is determined by the similarity between the tasks, which is little known and an on-going research problem in machine learning \cite{zamir2018taskonomy}. A straightforward way to resolve this issue is to collect enough additional labeled training samples to train a neural network for the new setting from scratch. This will achieve good performance, but it is impractical in real systems because training neural networks requires a large amount of samples and long computing time. To cope with the dynamics of wireless networks, it is highly desirable to reduce the training time for the new task, i.e., when the network setting changes. As discussed in Section III.F, LORM enjoys certain generalization capability, and can handle minor scenario changes. But its performance deteriorates when the setting changes dramatically. Note that although the tasks are different, they share something in common, i.e., the same structure of the underlying optimization problem. In other words, the knowledge learned in the old task can be helpful for the new task. Thus, it is possible to train a new model with only a few additional samples if we can effectively transfer such knowledge. Such an approach is called \emph{transfer learning}, which can significantly reduce the training time and achieve good performance in the new task. The difference between traditional machine learning and transfer learning is shown in Fig. \ref{fig:tl}.

\vspace{-2em}
\subsection{Self-imitation with Unlabeled Samples} \label{sec:explore}
\setlength{\textfloatsep}{3pt plus 1.0pt minus 2.0pt}
\begin{figure*}[htb]
	\centering
	\includegraphics[width=0.7\textwidth]{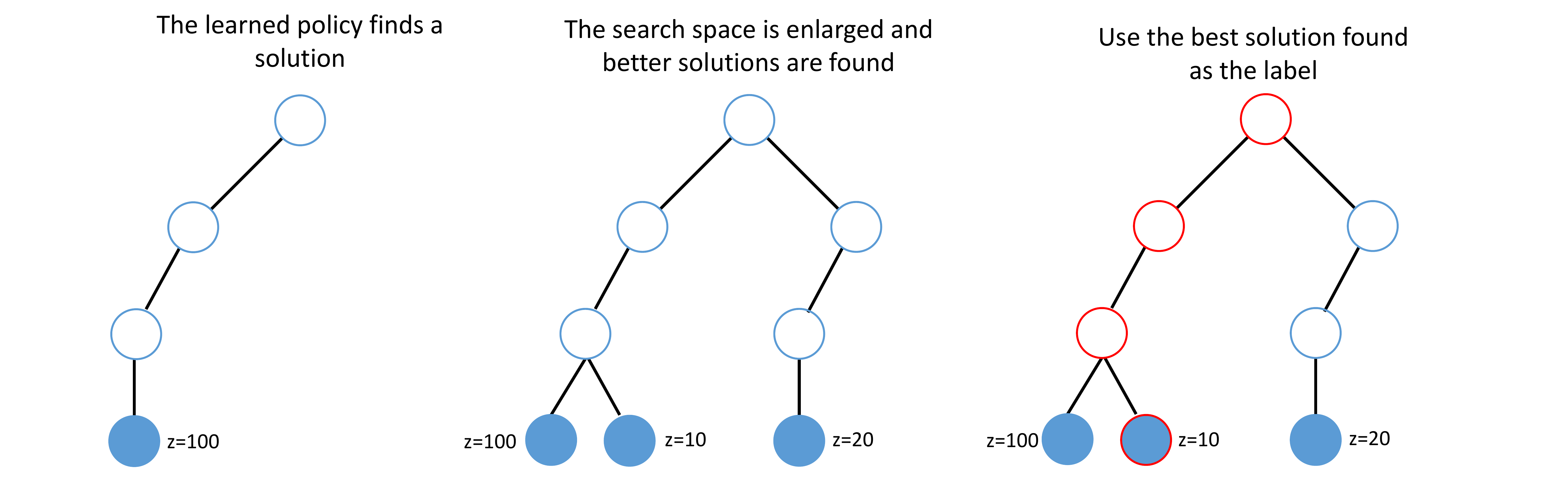}
	\caption{ An illustration of the self-exploration algorithm. The learned policy from the original task fails to find a good solution. Then, the search space is enlarged to find more solutions by tuning the threshold $\Lambda$. The best solution found is used as the label for LORM.}
	\label{fig:ro}
\end{figure*}
In this subsection, we first present transfer learning via fine-tuning and discuss its limitations. Then, we introduce an approach called \emph{self-imitation} to address the problems for fine-tuning. 

Fine-tuning is the most frequently employed method for transfer learning in neural networks. Neural networks are usually trained via stochastic gradient descent (SGD). For different layers, we can have different learning rates, i.e., the step size of the SGD. Fine-tuning is to tune the learning rate of each layer to refine the pre-trained neural network on the additional training dataset. We can train the pre-trained neural network with a small learning rate. The knowledge learned on the original dataset can serve as a good initialization point. The learning rate for fine-tuning is small because we expect that the initial weights are relatively good, so distorting them too much and too quickly is not a smart choice. The illustration of fine-tuning is shown in Fig. \ref{fig:ft}. Fine-tuning reduces the training time, but it needs additional labeled samples. The time cost is still expensive as the computational complexity of branch-and-bound to generate the labels, i.e., the optimal solutions, is exponential. This implies that we even have difficulty in generating a small amount of training labels if the network size of the new setting is large. Hence, it will be desirable if we can refine the model with unlabeled data.

\setlength{\textfloatsep}{3pt plus 1.0pt minus 2.0pt}
\begin{algorithm}  
	\caption{DAgger($\kappa$)}  
	\label{alg:SI}  
	\small
	\begin{algorithmic}[1]
		\State $\mathcal{D} \leftarrow \{\}, c^{*(j)} \leftarrow + \infty, \mathcal{N}_{opt}^{(j)} \leftarrow \{\}, \mathcal{T}^{(j)} \leftarrow \{\}, \pi^{(0)}\leftarrow\kappa$
		\For{$k \leftarrow 1,\cdots,M$ }
		\For{$j \leftarrow 1,\cdots,|\mathcal{P}|$ }
		\State Set a high threshold $\Lambda$ in $\pi^{(k-1)}$
		\State $c^{*(j)}, \mathcal{N}_{opt}^{(j)}, \mathcal{D}^{(kj)}, \mathcal{T}^{(j)}, \leftarrow \text{COLLECT-SI}(\pi^{(k-1)},\mathcal{P}_{j}, \mathcal{N}_{opt}^{(j)}, \mathcal{T}^{(j)}, c^{*(j)})$
		\State $\mathcal{D} \leftarrow \mathcal{D} \cup \mathcal{D}^{(kj)}$
		\EndFor
		\State $\pi^{(k)} \leftarrow$ fine-tune the classifier $\pi^{(k)}$ using data $\mathcal{D}$
		\EndFor
		\State \Return best $\pi^{(k)}$ on validation set
	\end{algorithmic}  
\end{algorithm}

\setlength{\textfloatsep}{3pt plus 1.0pt minus 2.0pt}
\begin{algorithm}  
	\caption{COLLECT ($\kappa$, p, $\mathcal{N}_{old}$, $\mathcal{T}$, $c^*_{old}$)}  
	\label{alg:COLLECT_SI} 
	\small
	\begin{algorithmic}[1]
		\State $\mathcal{L}\leftarrow \{{\rm {N}}_1\},\mathcal{D}\ \leftarrow \{\},t \leftarrow 0, \mathcal{N}_{opt} \leftarrow \{{\rm {N}}_1\}$
		\State Perform Algorithm \ref{alg:new_BnB} with $\kappa$
		\Statex $c^* \leftarrow$ best solution found, $\rm{N}_{opt} \leftarrow$ the node containing optimal solution
		\Statex Save all visited nodes and features into $\mathcal{D}$
		\Statex Save solutions of solved convex problems into $\mathcal{T}$
		
		\If {$c^* < c^*_{old}$}
		\State $\mathcal{N}_{opt} \leftarrow $ the nodes from root to $\rm{N}_{opt}$
		\Else 
		\State $\mathcal{N}_{opt} \leftarrow \mathcal{N}_{old}$, $c^* \leftarrow c^*_{old}$
		\EndIf

		\For{$j \leftarrow 1,\cdots,|\mathcal{D}|$ }
		\If{$\text{node}(D_j) \in \mathcal{N}_{opt}$}
		\State $\mathcal{D}_j \leftarrow \{\mathcal{D}_j.f,preserve\}$
		\Else
		\State $\mathcal{D}_j \leftarrow \{\mathcal{D}_j.f,prune\}$
		\EndIf
		\EndFor

		\State \Return $c^*$, $\mathcal{N}_{opt}$, $\mathcal{D}$, $\mathcal{T}$
		
	\end{algorithmic}  
\end{algorithm}

We propose to use transfer learning via self-imitation \cite{songlearning} to address the task mismatch problem with a few additional unlabeled samples. Self-imitation takes advantage of the exploration in reinforcement learning, which leads to an unsupervised paradigm. The key idea is to explore the search tree, and use the best solution found during the exploration as the label for imitation learning. In this way, the search space is enlarged and may contain better solutions than the learned policy. To transfer the knowledge in the learned policy, we harness the method proposed in Section \ref{sec:mlp} to enable a larger and better search space. This exploration policy is more effective than traditional exploration policies, e.g., $\epsilon$-greedy policy \cite{lavet2018introduction}.

The old policy $\pi^{(0)}$ was trained on the original training dataset. As the task changes, we collect a few additional unlabeled samples. The performance of the learned policy may not be good on the new training dataset. We set the learned policy with a higher threshold $\Lambda$ as the exploration policy. Algorithm \ref{alg:new_BnB} is performed on the new training dataset with the exploration policy. It will achieve better performance than the learned policy because the search space is larger. For each problem, we record the best solution found $c^*$ during the exploration process. The nodes on the path from the root to the one containing the best solution are marked as the optimal nodes, i.e., labeled as \emph{preserve}. Other nodes are labeled as \emph{prune}. A new policy $\pi^{(1)}$ is trained on the new training dataset. We repeat such a process with DAgger \cite{ross2011reduction}. The performance of the labels is improved iteratively by repeating such a process, and thus the performance of the learned policy will be improved iteratively. DAgger goes through the training dataset several times and we need to solve the convex problems at the leaf node each time. Thus, some convex problems are solved repeatedly in the training process. We propose to use the lookup table $\mathcal{T}$ proposed in Section \ref{sec:imp} to store the solutions. This will speed up the training process.

\paragraph*{Remark} The idea of self-imitation has been proposed and applied to solve medium-size MILPs in \cite{songlearning}. There are a few challenges to apply it to the LORM framework. Firstly, an incremental strategy was adopted in \cite{songlearning} to transfer to a different size. For example, in order to transfer from a network with $5$ users to a network with $15$ users, we must obtain training data of the networks with $5,6,\cdots,15$ users. It is impractical to get these data in wireless networks. Indeed, transferring from the network with $5$ users to the network with $15$ users directly will not provide a good performance because of the ineffectiveness of traditional exploration policies. Thus, we propose to increase the threshold $\Lambda$ of the pruning policy as the exploration policy. Such a policy increases the search space, which is the requirement of the exploration policy, and possesses the knowledge in the original task. In addition, since some relaxed problems are solved repeatedly during the training, we further use the lookup table proposed in Section \ref{sec:imp} to accelerate the training process.

\section{Simulations} \label{sec:simu}

In this section, we test the performance of LORM and LORM-TL, with the network power minimization problem in Cloud-RANs as an example. 

\subsection{Network Power Minimization in Cloud-RANs}\label{sec:sys_model}
The network power minimization problem in Cloud-RANs \cite{shi2014group} is a typical MINLP problem in wireless networks, as it consists of discrete variables (i.e., the selection of remote radio heads (RRHs) and fronthaul links), continuous variables (i.e., downlink beamforming coefficients), and QoS constraints and power constraints. 

Consider a Cloud-RAN with $L$ RRHs and $K$ single-antenna mobile users (MUs), where the $l$-th RRH is equipped with $N_l$ antennas. The baseband unit (BBU) pool is connected to all the RRHs via a high-bandwidth, low-latency fronthaul network, and performs centralized signal processing. The corresponding SINR for the $k$-th MU is given by $${\rm SINR}_k = \frac{|\sum_{l \in \mathcal{L}}\vct{h}_{kl}^{\sf{H}}\vct{w}_{lk}|^2}{\sum_{i\neq k}|\sum_{l \in \mathcal{L}}\vct{h}_{kl}^{\sf{H}}\vct{w}_{li}|^2 + \sigma_k^2}, \forall k \in \mathcal{S},$$ where $\vct{w}_{lk} \in \mathbb{C}^{N_l}$ denotes the transmit beamforming vector from RRH $l$ to MU $k$, $\vct{h}_{kl} \in \mathbb{C}^{N_l}$ denotes the channel vector between the $k$-th MU and the $l$-th RRH, and $\sigma_k^2$ is the variance of additive noise.

Let a binary vector $\vct{a}=(a_1,\cdots,a_L)$ with $a_i\in\{0,1\}$ denote the mode
of each RRH, i.e., $a_i = 1$ if the $i$-th RRH and the corresponding
transport link are switched on. Each RRH has its own transmit power constraint 
$$\sum_{k=1}^K \|\vct{w}_{lk}\|_{2}^2 \leq a_l \cdot P_l, l \in \{1,\cdots,L\},$$ 
where $\|\cdot\|_{2}$ is the $\ell_2$-norm of a vector.

The network power consumption in Cloud-RAN consists of the relative fronthaul network power consumption and the total transmit power consumption \cite{shi2014group}.  Assuming that the target SINR for the $k$-th user is $\gamma_k$, the network power minimization problem is formulated as the following MINLP problem:

\begin{equation}
\begin{aligned}
&\underset{\vct{w},\vct{a}}{\text{minimize}}
& & \sum_{l=1}^{L} a_l\cdot P_l^c + \sum_{l=1}^{L}\sum_{k=1}^{K} \frac{1}{\eta_l} \|\vct{w}_{k}\|_{2}^2\\
& \text{subject to}
& & \sqrt{\sum_{i \neq k}|\vct{h}_k^{\sf{H}}\vct{w}_i|^2 + \sigma_k^2 } \leq \frac{1}{\gamma_k}\Re(\vct{h}_k^{\sf{H}}\vct{w}_k), \forall k \\
& 
& &\sqrt{\sum_{k=1}^{K} \|\mtx{A}_{lk}\vct{w}_{k}\|_{2}^2} \leq a_l \cdot \sqrt{P_l}, \forall l \\
&
& & a_l \in \{0,1\}, \forall l,
\end{aligned}
\end{equation}
where $P_l^c$ is the relative fronthaul link power consumption, i.e., the power saved when both the RRH and the corresponding fronthaul link are switched off, and $\eta_l$ is the drain efficiency of the radio frequency power amplifier. Likewise, the aggregative beamforming vector is $\vct{w}=[\vct{w}_1^T,\dots, \vct{w}_K^T]^T\in\mathbb{C}^{NK}$ with $\vct{w}_k = [\vct{w}_{1k}^T, \dots, \vct{w}_{Lk}^T]^T \in \mathbb{C}^{N}$, $N=\sum_{l=1}^L N_l$, $\vct{h}_k = [\vct{h}_{1k}^T,\dots, \vct{h}_{Lk}^T]^T $ $\in \mathbb{C}^{N}$, $\Re(\cdot)$ denotes the real part of a complex scalar, and $\mtx{A}_{lk} \in \mathbb{C}^{N \times N}$ is a block diagonal matrix with an identity matrix $\mtx{I}_{N_l}$ as the $l$-th main diagonal block matrix and zeros elsewhere. This problem is reduced to a second-order cone programing (SOCP) problem if the integer variable $\vct{a}$ is fixed. Thus, it belongs to the problem class $\mathscr{P}$ and can be solved by LORM.
In our simulations, we consider the channel model $\vct{h}_{kl} = 10^{-L(d_{kl})/20}\sqrt{\phi_{kl} s_{kl}} \vct{g}_{kl}$, where $L(d_{kl})$ is the path-loss at distance $d_{kl}$, $\phi_{kl}$ is the antenna gain, $s_{kl}$ is the shadowing coefficient, and $\vct{g}_{kl}$ is the small scale fading coefficient. One can refer to Table I in \cite{shi2014group} for the detailed parameter settings. The RRHs and MUs are uniformly and independently distributed in the square region $[-1000, 1000] \times [-1000, 1000]$ meters, i.e., their locations are different in different samples. The fronthaul link power consumption is set to be a random permutation of $P^c_l = (5+l)W, l=1, \cdots, L$ in different samples.

\subsection{Performance Comparison}
We first compare the proposed methods with the state-of-the-art methods designed for the network power minimization problem in Cloud-RANs, in terms of the performance and computation time. For LORM and LORM-TL, we set the hidden units of the neural network as $\{32,64,16\}$. We use the branching variable's value, i.e., $a_i[j]$, and its value at the root problem, i.e., $a_0[j]$, as problem-independent features. Motivated by the observations in \cite{shi2014group}, we strive to open a small number of RRHs with a small fronthaul link power consumption if the problem is feasible. Thus, the problem-dependent features are selected as the fronthual link power consumption, i.e., $\frac{P^c_j \cdot L}{\sum_{k=1}^L P^c_k}$. In addition, we use the iterative algorithm to guarantee the feasibility, as introduced in Section \ref{sec:mlp}, and set the threshold $\Lambda$ as $1-0.5\cdot 0.8^l$, where $l$ is the iteration index. We set the number of DAgger as $M=5$, and train $5$ epochs for each DAgger iteration.

The following three benchmarks are considered:
\begin{enumerate}
	\item Relaxed Mixed-integer Nonlinear Programming based algorithm (RMINLP) \cite{cheng2013joint}: This algorithm is a commonly used heuristic algorithm for MINLPs. It applies a deflation procedure to switch off RRHs one-by-one based on the solution of a relaxed problem. It needs to solve $3L$ SOCPs.
	
	\item Iterative Group Sparse Beamforming (GSBF) \cite{shi2014group}: This algorithm is designed for the network power minimization problem in Cloud-RANs, and is the state-of-the-art solutions for this problem. It needs to solve $L$ SOCPs.
	
	\item Branch-and-Bound algorithm (Optimal solution): This is the standard branch-and-bound algorithm with the best-first node selection policy.
\end{enumerate}

\begin{figure}[htb]
	\centering
	\includegraphics[width=0.5\textwidth]{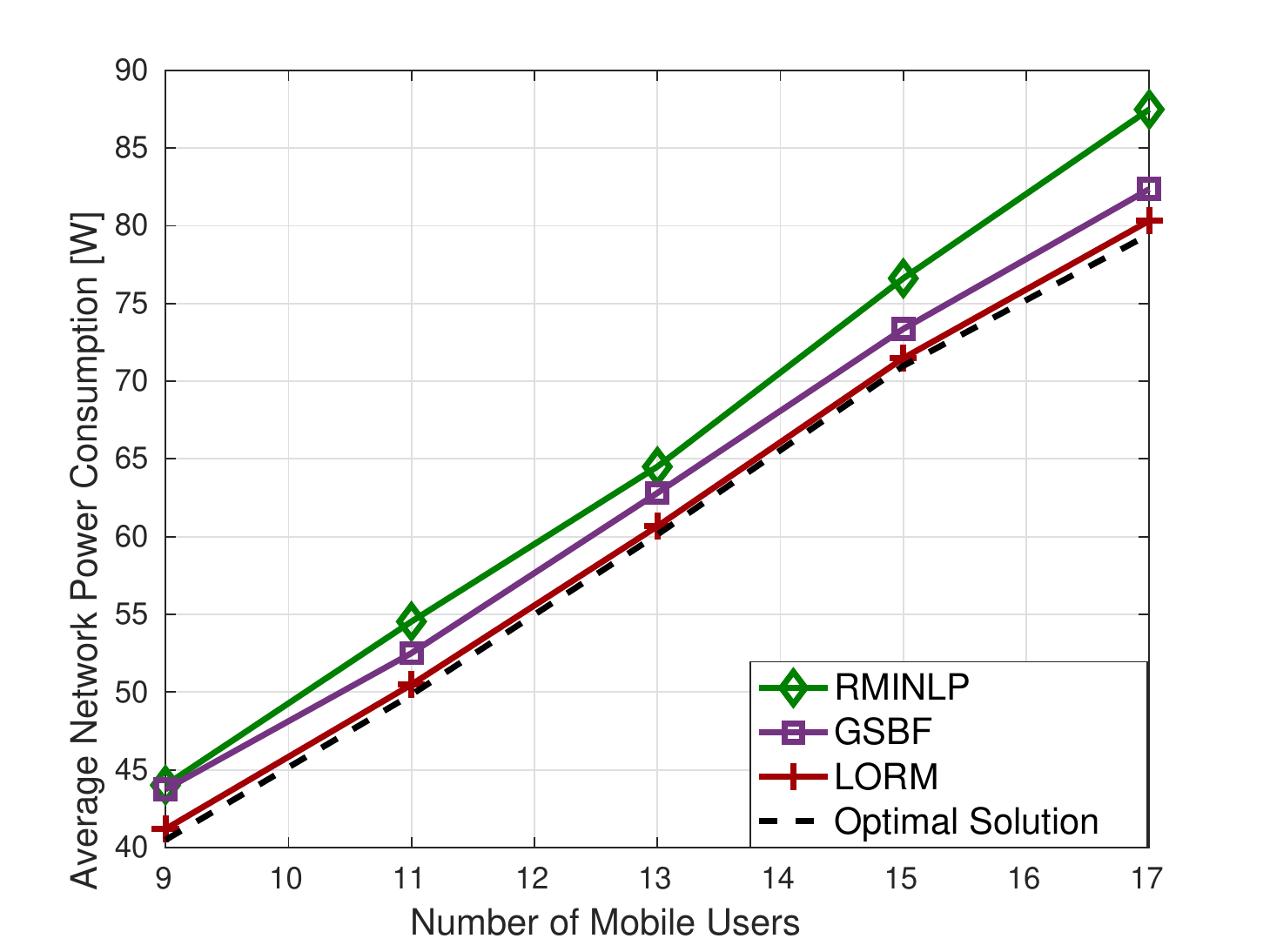}
	\caption{The average network power consumption versus the number of mobile users.}
	\label{fig:per1}
\end{figure}

\subsubsection{Performance of LORM} We simulate the performance under the setting with the target SINR (TSINR) as $4$dB, with $10$ 2-antenna RRHs and different numbers of users. For each setting, we generate $50$ network realizations, i.e., samples, for training and $50$ network realizations for testing. The performance is shown in Fig. \ref{fig:per1}, and the computation time is shown in Table \ref{tab:time_comp}. It is observed that LORM outperforms both GSBF and RMINLP, and achieves near-optimal performance. We also see that LORM has a much less running time than RMINLP and branch-and-bound, and achieves about 2x speedup compared with GSBF. The locations of RRHs and MUs in the training dataset and test dataset are different, and thus the results already demonstrate a certain generalization capability of LORM, i.e., generalization to different large-scale fading conditions.

\setlength{\textfloatsep}{3pt plus 1.0pt minus 1.5pt}
\begin{table}[htb]
	
	\selectfont  
	\centering
	\newcommand{\tabincell}[2]{\begin{tabular}{@{}#1@{}}#2\end{tabular}}
	\caption{Average running time for the algorithms under each setting (in seconds).} 
	
	\resizebox{0.7\textwidth}{!}{
		\begin{tabular}{|c|c|c|c|c|c|c|}  
			\hline  
			Setting  & Branch-and-Bound &  LORM & GSBF & RMINLP \cr\hline
			$L=10$, $K=13$, $\text{TSINR}=4$ &91.76s&0.892s&2.562s&5.264s\cr\hline
			$L=10$, $K=15$, $\text{TSINR}=4$ &96.40s&1.157s&3.680s&8.136s\cr\hline 
			$L=10$, $K=17$, $\text{TSINR}=4$ &142.0s&2.920s&5.474s&12.64s\cr\hline   
			
	\end{tabular}}
	\label{tab:time_comp}
\end{table}

\subsubsection{Generalization of LORM} We conduct the following experiment to test the generalization capability of LORM beyond different large-scale fading. We first train the model on $100$ samples with $L=6$, $K=9$, $\text{TSINR}=0$, and test it on $L=10$, $\text{TSINR}=4$, with different numbers of users. The results are shown in Table \ref{tab:bf_gen}, where ``full training'' means training with sufficient data for that setting. From the table, we see that even if trained on a different setting, LORM outperforms both GSBF and RMINLP, thus demonstrating its good generalization ability. 

\setlength{\textfloatsep}{3pt plus 1.0pt minus 2.0pt}
\begin{table}[htb]
	
	\selectfont  
	\centering
	\newcommand{\tabincell}[2]{\begin{tabular}{@{}#1@{}}#2\end{tabular}}
	\caption{The gap from the optimal objective value.} 
	
	\resizebox{0.8\textwidth}{!}{
		\begin{tabular}{|c|c|c|c|c|c|c|}  
			\hline  
			Setting  & \tabincell{c}{LORM \\ (full training)} & \tabincell{c}{LORM \\ (train on $L=6$, $K=9$, $\text{TSINR}=0$)} & RMINLP & GSBF \cr\hline
			$L=10$, $K=7$, $\text{TSINR}=4$ & $1.87\%$& $4.39\%$    &$7.94\%$    &  $12.9\%$   \cr\hline
			$L=10$, $K=9$, $\text{TSINR}=4$ &$1.73\%$&$2.97\%$&$8.71\%$&$8.04\%$\cr\hline 
			$L=10$, $K=11$,$\text{TSINR}=4$ &$1.30\%$&$1.72\%$ &$9.44\%$&$5.36\%$\cr\hline 
			$L=10$, $K=13$, $\text{TSINR}=4$ &$1.41\%$&$1.41\%$&$7.27\%$&$4.45\%$\cr\hline
			$L=10$, $K=15$,$\text{TSINR}=4$ &$0.70\%$&$1.48\%$&$7.94\%$&$3.36\%$\cr\hline 
			
	\end{tabular}}
	\label{tab:bf_gen}
\end{table}

\subsubsection{Performance of LORM-TL} Throughout experiments, we observed performance deterioration when the number of MUs change dramatically. In this part, we test the effectiveness of LORM-TL in this challenging situation, where LORM cannot achieve good performance. We conduct two experiments: When the number of MUs in the test dataset is either much less or much larger than the training dataset. In the first experiment, we train LORM-TL on $100$ labeled samples with $L=6$, $K=6$, $\text{TSINR}=0$. We then transfer on $10$ unlabeled samples with $L=10$, $K=15$, and test on $L=10$, $K=15$. The results are shown in Table \ref{tab:bf_si1}. In the second experiment, we train LORM-TL on $50$ labeled samples with $L=10$, $K=17$, $\text{TSINR}=4$, and then transfer on $10$ unlabeled samples with $L=10$, $K=7$, and test on $L=10$, $K=7$. For the self-imitation algorithm, the number of iterations for self-imitation is set as $M=10$. The threshold $\Lambda$ is set as $0.9$ at the first iteration, and is halved if this iteration's performance is the same as the last iteration. The results are shown in Table \ref{tab:bf_si2}. As shown in the tables, equipped with transfer learning, LORM-TL outperforms all competing methods, and achieves comparable performance with LORM trained on sufficient samples of the same distribution with the test dataset.

\begin{table}[htb]
	
	\selectfont  
	\centering
	\newcommand{\tabincell}[2]{\begin{tabular}{@{}#1@{}}#2\end{tabular}}
	\caption{The performance gap of each method to the optimal objective value.} 
	
	\resizebox{0.6\textwidth}{!}{
	\begin{tabular}{|c|c|c|c|c|c|c|}  
		\hline  
		Setting &\tabincell{c}{LORM\\(train on $L=6$, \\ $K=6$, $\text{TSINR}=0$)}& \tabincell{c}{LORM-TL} &  \tabincell{c}{LORM\\(full training)}& RMINLP & GSBF \cr\hline
		\tabincell{c}{$L=10$, $K=15$\\$\text{TSINR}=0$} &--\tablefootnote{To control the time during the test, we set the maximum number of iterations as $5$. For the method in this column, it only finds feasible solutions for $82\%$ of the problems during the test. Hence, we do not record its performance here.}&$0.63\%$&$0.61\%$&$7.07\%$&$6.76\%$\cr\hline 
		\tabincell{c}{$L=10$, $K=15$\\$\text{TSINR}=2$} &--&$0.38\%$&$0.3\%$&$7.92\%$&$5.06\%$\cr\hline 
		\tabincell{c}{$L=10$, $K=15$\\$\text{TSINR}=4$} &--&$0.68\%$&$0.39\%$&$7.94\%$&$3.36\%$\cr\hline
		
\end{tabular}}
	\label{tab:bf_si1}
\end{table}

\begin{table}[htb]
	
	\selectfont  
	\centering
	\newcommand{\tabincell}[2]{\begin{tabular}{@{}#1@{}}#2\end{tabular}}
	\caption{The performance gap of each method to the optimal objective value.} 
	
	\resizebox{0.6\textwidth}{!}{
		\begin{tabular}{|c|c|c|c|c|c|c|}  
			\hline  
			Setting &\tabincell{c}{LORM\\(train on $L=10$, \\ $K=17$, $\text{TSINR}=4$)}&\tabincell{c}{LORM\\(full training)}& \tabincell{c}{LORM-TL} &  RMINLP & GSBF \cr\hline
			\tabincell{c}{$L=10$, $K=7$\\$\text{TSINR}=0$} &$57.1\%$&$0.72\%$&$2.48\%$&$7.43\%$&$10.0\%$\cr\hline 
			\tabincell{c}{$L=10$, $K=7$\\$\text{TSINR}=2$} &$43.8\%$&$2.30\%$&$2.79\%$&$5.57\%$&$11.9\%$\cr\hline 
			\tabincell{c}{$L=10$, $K=7$\\$\text{TSINR}=4$} &$29.9\%$&$1.87\%$&$2.95\%$&$6.89\%$&$12.9\%$\cr\hline
			
	\end{tabular}}
	\label{tab:bf_si2}
\end{table}

\vspace{-2em}
\subsection{Comparison with Existing ``Learning to Optimize'' Approaches}
In this part, we further compare with other learning-based methods to show the effectiveness and efficiency of LORM. The main method we compare with is a modified version of \cite{he2014learning}, which is referred to as ``Modified MILP-SVM''. The original method in \cite{he2014learning} is proposed for MILPs, and it can be applied to MINLPs with minor modifications. Specifically, we remove some features that are available only for MILPs and add problem-dependent features. Furthermore, since the method in \cite{he2014learning} has difficulty to guarantee feasibility, the benchmark method we implemented is tuned on the test dataset instead of the validation dataset for obtaining a feasible solution. We also implemented the method in \cite{sun2018learning}, trained on $20,000$ samples generated by the state-of-the-art heuristic algorithm \cite{shi2014group}, but the output solutions were not feasible. Therefore, we do not include it in the comparison. 

The performance gap to the optimal solution and the computation time are recorded under the settings shown in Table \ref{tab:comp_nips_per}. For each setting, we generate $50$ network realizations, i.e., samples, for training and $50$ network realizations for test. From the table, we see that MILP-SVM also accelerates the branch-and-bound for MINLP by tuning on the test dataset to guarantee the feasibility. Nevertheless, the computation burden of MILP-SVM is still high. It is shown that LORM achieves comparable performance with 10x speedup to MILP-SVM. This demonstrates the efficiency of LORM in MINLP problems, and verifies the computational complexity analysis in Section \ref{sec:ana}. 


\setlength{\textfloatsep}{3pt plus 1.0pt minus 2.0pt}
\begin{table}[htb]
	
	\selectfont  
	\centering
	\newcommand{\tabincell}[2]{\begin{tabular}{@{}#1@{}}#2\end{tabular}}
	\caption{The performance gap to the optimal objective value and the average computation time of each method.} 
	
	\resizebox{0.9\textwidth}{!}{
		\begin{tabular}{|c|c|c||c|c|c|c|}  
			\hline 
			& \multicolumn{2}{c||}{Performance Gap} & \multicolumn{3}{c|}{Time}        \cr\hline
			Settings & LORM      & Modified MILP-SVM      & LORM & Modified MILP-SVM & Branch-and-Bound\cr\hline
			$L=6$, $K=8$, TSINR=0 & $0.02\%$& $0.02\%$& 0.102s     &  0.68s  & 3.24s\cr\hline
			$L=6$, $K=8$, TSINR=2 & $1.54\%$&$1.63\%$& 0.131s  & 0.75s & 3.1s               \cr\hline
			$L=6$, $K=8$, TSINR=4 & $0.80\%$&$0.62\%$& 0.118s &1.098s & 2.9s                        \cr\hline
			$L=10$, $K=15$, TSINR=0 & $0.35\%$&$0.63\%$ & 0.974s &6.634s & 107.2s     \cr\hline
			$L=10$, $K=15$, TSINR=2 & $0.59\%$&$0.32\%$& 1.09s&10.08s & 114.2s \cr\hline
			$L=10$, $K=15$, TSINR=4 & $1.25\%$&$1.17\%$& 1.1s&11.04s & 96.4s\cr\hline
	\end{tabular}}
	\label{tab:comp_nips_per}
\end{table}

\section{Conclusions}
In this paper, we proposed two machine learning-based frameworks, namely LORM and LORM-TL, to achieve near-optimal performance for MINLP resource management problems in wireless networks. We first proposed LORM, an imitation learning-based approach, to learn the pruning policy in the optimal branch-and-bound algorithm. By exploiting the algorithm structure, LORM not only achieves near-optimal performance with few training samples, but also guarantees feasibility of the constraints, and can be generalized to problem instances with larger sizes than the training dataset. Furthermore, we proposed LORM-TL to address the task mismatch issue for LORM by relying on only a few additional unlabeled training samples. Extensive experiments have been conducted to verify the effectiveness of both LORM and LORM-TL.

Different from previous approaches for ``learning to optimize'' that directly approximate the input-output mapping, our proposed framework achieves few-shot learning by exploiting the algorithm structure. As training samples are in general expensive to obtain, we expect that the proposed framework will facilitate the adoption of machine learning-based methods in wireless communications. For the next stage, it will be interesting to test the proposed methods on other design problems in wireless networks, and develop approaches to further reduce the sample complexity and computational complexity.
\vspace{-1em}
\appendices 
\section{Proof of Theorem \ref{thm:socp}}
We first derive the expected number of nodes explored. Consider an $n$-layer full binary tree. Denote the number of nodes whose root is an optimal node as $a(n)$, and the number of nodes whose root is not an optimal node as $b(n)$. Further assume that the node pruning policy expands a non-optimal node with probability $\epsilon_1$ and prunes an optimal node with probability $\epsilon_2$. Define $E_1$ as the expected number of nodes explored and $E_2$ as the number of relaxed problems to solve.

Then, we have the following recurrence formula
\begin{equation}\label{eq:rec_1}
\left\{
\begin{aligned}
&a(n) = (1-\epsilon_2)\cdot a(n)+\epsilon_1\cdot b(n) + 1 \\
&b(n) = 2\epsilon_1\cdot b(n-1)+1, \\
\end{aligned}
\right.
\end{equation}
with initial states as $a(1)=b(1)=1.$

The number $a(n)$ is needed for computing the number of nodes explored. Solving the recurrence formula, we obtain
\begin{equation}\label{eq:a_n1}
a(n)=\left\{
\begin{aligned}
&(1-\epsilon_2)^{n-1}(1-\sigma_2-\sigma_1)+\sigma_2\cdot(2\epsilon_1)^{n-1}+\sigma_1, &\epsilon_2 \neq 0, \\
&n + \frac{\epsilon_1}{1-2\epsilon_1}\left(n-1-\frac{2\epsilon_1}{1-2\epsilon_1}\right) + \frac{\epsilon_1 (2\epsilon_1)^{n}}{(1-2\epsilon_1)^2}, &\epsilon_2=0,\\
\end{aligned}
\right.
\end{equation}
where $\sigma_1=\frac{1-\epsilon_1}{\epsilon_2(1-2\epsilon_1)}$, $\sigma_2=\frac{2\epsilon_1^2}{(1- 2\epsilon_1 - \epsilon_2)(1-2\epsilon_1)}$.

Taking $E_1=a(L+1)$ completes the proof for the expected number of nodes. It is shown in (\ref{eq:rec_1}) that $a(n)$ is monotonically increasing when $\epsilon_1$ is increasing or $\epsilon_2$ is decreasing. Thus, in order to prove the computational complexity bound, it is sufficient to show that $E_2 = \mathcal{O}(L^2)$, when $\epsilon_1 = 0.5$ and $\epsilon_2 = 0$. By letting $\epsilon_1 = 0.5$ in (\ref{eq:a_n1}), we get $a(n) = 0.5(n-1)^2 +n.$
Therefore, $E_1 = a(L+1) = \mathcal{O}(L^2)$, when $\epsilon_1 \leq 0.5$. Similarly, taking $\epsilon_1 = 0.3$ and $\epsilon_2 = 1$, we get
$a(n) = 2n - 3 + 3\left(\frac{2}{3}\right)^n = \mathcal{O}(n).$ Therefore, $E_1 = a(L+1) = \mathcal{O}(L)$, when $\epsilon_1 \leq 0.3$. This completes the proof for the expected number of nodes.

Next, we consider the number of relaxed problems to be solved. Consider an $n$-layer full binary tree. Also denote the number of relaxed problems whose root is an optimal node as $c(n)$, and the number of relaxed problems whose root is not an optimal node as $d(n)$. Since we only solve the convex problems at the leaf nodes, the recurrence formula can be represented as 

\begin{equation}\label{eq:par}
\left\{
\begin{aligned}
&c(n) = (1-\epsilon_2)c(n-1) + \epsilon_1 d(n-1) \\
&d(n) = 2\epsilon_1 d(n-1) \\
\end{aligned}
\right.
\end{equation}
with initial states as $c(1) = d(1) = 1$.

Solving the recurrence formula, we get $c(n) \leq 1 + \epsilon_1\left(\frac{1-(2\epsilon_1)^{n-1}}{1-2\epsilon_1}\right)$ when $\epsilon_2 \leq 1, n\geq 2$. This implies $E_2 = c(L+1) \leq 1 + \epsilon_1 L = \mathcal{O}(L)$ when $\epsilon_1 \leq 0.5$, and $E_2 = a(L+1) \leq 3 = \mathcal{O}(1)$ when $\epsilon_1 \leq \frac{1}{3}$.

\vspace{-1em}
\bibliographystyle{ieeetr}
\bibliography{Reference}

\end{document}